Title: Off-center impurity in alkali halides: reorientation, electric polarization and pairing to F center. III. Numerical calculations.

Authors: G. Baldacchini (1), R.M. Montereali (1), U.M. Grassano (2), A. Scacco (3), P. Petrova (4), M. Mladenova (5), M. Ivanovich (6), and M. Georgiev (6) (ENEA C.R.E. Frascati, Frascati (Roma), Italy (1), Dipartimento di Fisica, Universita "Tor Vergata", Rome, Italy (2), Dipartimento di Fisica, Universita "La Sapienza", Rome, Italy (3), Institute of Nuclear Research and Nuclear Energy, Bulgarian Academy of Sciences, Sofia, Bulgaria (4), Department of Condensed Matter Physics, Faculty of Physics, University of Sofia, Sofia, Bulgaria (5), Institute of Solid State Physics, Bulgarian Academy of Sciences, Sofia, Bulgaria (6))

We carry out numerical calculations by an extended Hückel program in order to check the analytical results reported in the preceding papers I and II. We typically consider alkali halide clusters composed of some tens of constituent atoms to calculate electronic energies under static conditions or versus the displacements of particular atoms. Among other things, the off-center displacement of the substitutional $Li^+$ impurity in most alkali halide lattices is evidenced. The sinusoidal profile of the rotational barriers is also confirmed for KCl.

## 1. Introduction

In the preceding Parts I [1] and II [2] we derive analytically off-center displacements and barriers for the reorientational motion of a $Li^+$ impurity upon the off-center elipsoid. In this part III, we compare our analytical conclusions with numerical calculations using an extended Hückel sofware. We mostly calculate energies of electronic states of alkali halide clusters composed of some tens of constituent atoms under static conditions or under the displacements of specific atoms along certain crystallographic axes. We study both cation and anion displacements in two-and three-dimensional clusters.

In particular, we find the total electronic energy of a cluster centered at a $Li^+$ impurity minimal as the impurity is displaced away from the cation site, along <111> in a non-F- centered cluster or along <110> in the immediate (110) plane just beneath an F center. The off-center displacements derived from the minima are in concert with analytical calculations within an accuracy reasonable for the method. Calculated off-center displacements are summarized in Table I. Another impressing finding is the sinusoidal profile of the reorientational barriers as the $Li^+$ impurity is rotated along the off-center ring in the (110) plane just beneath an F center.

The problem of a small-size impurity embedded in a crystalline environment has been addressed earlier in a number of numerical calculations. In particular, we refer to a computer simulation study of off-center impurity dipoles in alkali halides and alkali earth oxides, and to references therein [3]. Off-center $Li^+$ displacements the order of one tenth of the interionic separation have been calculated for KF, KCl, and RbCl along <100>, <110> and <111>. We note similarities and dissimilarities between off- center displacements from various sources in Table I.

## 2. Numerical calculations

### 2.1. Preliminaries

We carried out numerical calculations using a recent CACAO (Computer Aided Composition of Atomic Orbitals) software product based on the Extended Hückel Method [4]. The program apparently yielded electronic energies accurate to within 10 percent which was suitable for the purpose.

Before all, we reproduced the details of an early calculation [5,6] of the energies of a cluster composed of three Cl- pairs nearest-neighboring a $Li^+$ ion in KCl and found them generally in concert. In particular, we found a pair of two opposite-parity states of symmetries $a_{1g}$ and $t_{1u}$, respectively, to split at ~ 7 eV when $Li^+$ was on-center. (This gap is nevertheless nearly twice as large as the one reported by [5].) Subjecting $Li^+$ to small displacements along <111> we obtained minima and maxima in $t_{1u}$ and $a_{1g}$, respectively, which reproduced to a reasonable extent at small Q the square-root parts of the adiabatic energies in equation (I.15). Fitting to eq. (I.15), we roughly estimated b ~ 20 eV/Å at $E_{\alpha\beta}$ ~ 7 eV.

### 2.2. Cation displacements

For extending the calculation, we constructed various $Li^+$-containing clusters of the *fcc* KCl lattice (Fig.1). We imitated off-center displacements Q of the $Li^+$ impurity by moving it stepwise away from the center, along <111> in a otherwise perfect cluster and along <110> in a cluster containing an electronically charged chlorine vacancy at [001] with $Li^+$ at [000], to model an $F_A$ (Li) center.

To be precise, the program automatically distributed the extra charge over the nearest-neighboring $K^+$ ions around the vacancy at Q = 0. However, it also tended to push that charge away to the outer-edge cations terminating a cluster, built symmetrically of $K^+$ and $Cl^-$ ions, as the displacement Q increased. We found that the extra charge could be kept around the vacancy if extra anions were added to at edge sites to border the cluster so as to oppose the pushing trends.

We found the total electronic energy of a cluster decreasing against Q in either case before reaching a minimum and rising again, as shown in Fig.2 and Fig.3. In a non-F-centered cluster, a <111> minimum appeared at 30 % of $a\sqrt{3}$, where $a$ = 3.147 Å was the anion-to-cation separation in KCl. This is to be compared with the three $Cl^-$-pair cluster, where a similar minimum of the total energy appeared at the same displacement (cf. Fig.2 and Table II).

The effect of a nearby $Cl^-$ vacancy led to the appearence of <110> minima at 50 % of $a\sqrt{2}$ and at 33 % of $a\sqrt{2}$ with the vacancy empty or containing an extra electron, respectively (see Fig.3 and Table II). While the latter cluster was regarded as one representing the $F_A$ center, the former was assumed to model an ionized $F_A$ center. Concomittantly, the F center electron was seen to depress the off-center radius of a nearby small-size impurity, confirming our previous conclusion [7]. Substituting $K^+$ for $Li^+$ to imitate an impurity-free F/α-centered crystal resulted in reshaping the potential energy diagram, the cluster now being destabilized by <110> $K^+$ displacements (Fig.3 and Table II). The off-center displacements thus appeared intrinsic to the impurity rather than to program artifacts.

We next simulated two cases of in-plane reorientation of an off-centered $Li^+$ impurity by rotating it stepwise around the <001> direction: one in a non- F centered cluster and the other in a cluster containing an F center, respectively. In the former, the rotation was in the $Li^+$ -containing plane parallel to the (001) plane. Rotational radius was the projection of the off- center radius onto this plane, as shown in Fig.4. In the latter, the $Li^+$ impurity was rotated in the (001) plane with radius now given by its net off-center radius, as illustrated in Fig.5. The rotational angle φ was measured from the <110> axis. We found the total electronic energy changing proportionaly to cos(4φ) in both cases, as shown in the Figures, confirming our present analytic conclusions (cf. equation (II.1)).

As we were dealing with small 46 element clusters in which only one single $Li^+$ impurity displacement was varied stepwise, our obtained off-center radii $r_d$ appeared overestimated (cf. Table I). We believe future calculations should take into account more of the associated lattice distortions.

### 2.3. Anion displacements

For simulating the anion motion, we constructed various two- or three- dimensional clusters (from 24 to 46 elements) of the *fcc* KCl lattice. In most cases, the clusters contained a $Li^+$ impurity but sometimes $Na^+$-containing or impurity-free clusters were considered. In all cases, an empty (α center) or electron-containing (F center) chlorine vacancy was present at [001], while the impurity (if any) was at [000]. We imitated the vacancy displacement as a stepwise halogen approach towards it.

#### 2.3.1. Two-dimensional clusters

We considered 24-element two-dimensional KCl clusters containing either an α or F center at [010] with either a $Li^+$ or $Na^+$ impurity at [000]. The [100] $Cl^-$ ion was displaced along <$\underline{1}$10> towards the vacancy.

We found the total electronic energy of an α−centered cluster decreasing against halogen displacements Q in either case ($Li^+$ or $Na^+$) before reaching a minimum at 50 % $a\sqrt{2}$, where a = 3.147 Å (see Fig.6 and Table III).

Adding an extra electron to model the $F_A$ center changed the total energy curve shape drastically (Fig.7). In this case the halogen displacement led to the total energy increasing up to 50 % $a\sqrt{2}$ (half the initial separation between the F center and the displaced $Cl^-$ ion). The total energy change was $\delta\Sigma E$ = 0.83 eV for $F_A$ (Li) and $\delta\Sigma E$ = 0.95 eV for $F_A$ (Na), in some agreement with experimental data [8] (cf. Table III).

### 2.3.2. Three-dimensional clusters

The three-dimensional configuration provided two nonindentical options for anion motion towards the vacancy:

(i) along $<\underline{1}01>$ direction in (010) (XZ) plane,

(ii) along $<\underline{1}\underline{1}0>$ direction in (001) (XY) plane.

The two motions were quantitatively distinguishable, particularly when the vacancy was empty (α center). We found the total electronic energy of a cluster decreasing against Q in both cases before reaching a minimum and rising again (see Fig.8).

In case (i) the off-center radius was $r_d = 0.71a$ (50 % $a\sqrt{2}$) and $\delta\Sigma E$ = 0.27 eV. The stable configuration predicted the appearance of two semi-vacancies centered symmetrically at two neighboring anion sites, [100] and [001], with the $Li^+$ ion resting at [000].

In case (ii), the effect of an α center in proximity on the $Cl^-$ displacement was apparently predominating. We found a small minimized displacement towards the vacancy with $r_d = 0.14a$ and a total energy change $\delta\Sigma E$ = 0.012 eV. (The $Li^+$ impurity was at [000].) A similar calculation of an impurity-free cluster (i.e. KCl: α cluster) made almost no difference, with $r_d = 0.14a$, $\delta\Sigma E$ = 0.016 eV (cf. Fig.8 - curve *c* and *b* and Table III). The conclusion was that the impurity did not affect the anion displacement in (ii).

We also investigated the off-center behaviour of the anion in the presence of an $F_A$ center in both (i) and (ii). The effect of the F center electron resulted in a drastic change of the total electronic energy curve shape: now the energy increased significantly against the anion displacement Q up to Q = $0.71a$, with rotational barriers $\delta\Sigma E$ = 0.79 eV for (i) and $\delta\Sigma E$ = 0.91 eV for (ii) (cf. Fig. 9 and Table III).

We finally simulated the rotation around <010> of a $Li^+$- neighbouring $Cl^-$ ion. That is, a $Cl^-$ initialy at [100] was rotated stepwise around a $Li^+$- at [000] with a radius equal to *a,* the anion-to-cation separation in KCl, there being an α or F center at [001]. The rotation angle φ was measured from the <100> direction (Fig. 10 and Fig.11). We found that the total electronic energy change could be described as $\sin^3(2\varphi)$ for both α and F-centered clusters, with rotational barriers $\delta\Sigma E$ = 0.5 eV for the former and $\delta\Sigma E$ = 1.3 eV for the latter (see Fig. 10 and Fig. 11 and Table III).

## 3. Conclusion

We carried out extended Hückel cluster calculations of alkali halide geometries to check the stability of defect containing structures against specific displacements of constituent atoms. In particular, the calculations confirmed to a resonable accuracy off-center displacements and reorientational barriers of cationic substitutional $Li^+$- impurities, either isolated or near F centers. We believe these results form the basis for future more elaborate computer experiments on $F_A$ centers.

## Table I

Off-center $Li^+$ impurity radius $r_d$ in KCl

| $Li^+$ <111> displacement isolated $Li^+$ cluster | $Li^+$ <110> displacement $F_A(Li^+)$ cluster | $\alpha_A(Li^+)$ cluster |
|---|---|---|
| 0.52 * | 0.46 * | 0.71 * |
| 0.35 [a] | 0.21 [e] | |
| 0.36 [b] | 0.04 [f] | |
| 0.12 [c] | 0.29 [g] | |
| 0.205 [d] | | |

* $r_d$ is the minimized displacement along <111> and <110> in units of the anion-to-cation separation *a* = 3.147 Å. Calculated data: *present; [a]Van Winsum, J.A. *et al.* J. Phys. Chem. Solids **39**, 1217 (1978); [b]Glinchuk, M.D. in: "The Dynamical Jahn-Teller Effect in Localized Systems.", ed. Yu.E. Perlin and M. Wagner, Elsevier Science Publishers BV

(1984); [c]Fitzsimons, P.B. *et al*. Crystal Latt. Def. & Amorphous Mater. **15**, 7 (1987); [d]Catlow, C.R.A. *et al*. Phys. Rev. B **18**, 2739 (1978); [e]Sangster. M.J.L., J. Phys. C: Solid State Phys. **19**, 1107 (1986). Experimental data: Luty, F. Surf. Sci. **37**, 1120 (1973); May, M. *et al*., Optics Commun. **61**, 325 (1987).

Table II

Numerical calculations of $Li^+$ off-center displacements in KCl

| 3-D case | extremum type $r_d$ (Å) | extremum type $\delta_{EE}$ (eV) | | extremum type $r_d$ (Å) | extremum type $\delta_{EE}$ (eV) |
|---|---|---|---|---|---|
| $Li^+$ motion along <111> | | | <110> $K^+$ steps impurity free | | |
| vacancy free | minimum | minimum | cluster | maximum | maximum |
| large cluster | 0.52 | 0.30 | α centered | 0.71 | 0.24 |
| small cluster | 0.52 | 0.22 | F centered | 0.71 | 0.33 |
| $Li^+$ motion along <110> | minimum | | off-center $Li^+$ rotation | minimum | |
| α centered | 0.71 | 0.52 | vacancy free | | 0.14 |
| cluster | | | α centered | | 1.17 |
| F centered | 0.46 | 0.27 | F centered cluster | | 0.28 |

* $r_d$ is the magnitude of off-center displacement in units of the anion-to-cation separation of the KCl lattice, *a* = 3.147 Å. $\delta_{EE}$ is the potential energy barrier for anion motion.

Table III

Numerical calculations of Cl- extremal displacements in KCl*

| 2-D case | extremum $r_d$ (Å) | type $\delta_{EE}$ (eV) | | extremum $r_d$ (Å) | type $\delta_{EE}$ (eV) |
|---|---|---|---|---|---|
| $Cl^-$ motion | | | α centered | | |
| along <110> | | | impurity free cluster<br>(i) or (ii) | minimum<br>0.14 | minimum<br>0.016 |
| α centered cluster<br>$Li^+$ containing<br>$Na^+$ containing | minimum<br>0.71<br>0.71 | minimum<br>0.56<br>0.32 | | | |
| F centered cluster<br>$Li^+$ containing<br>$Na^+$ containing | maximum<br>0.71<br>0.71 | maximum<br>0.83<br>0.95 | F centered $Li^+$ cluster<br>(ii)-<101><br>(ii)-<110> | maximum<br><br>0.71<br>0.71 | maximum<br><br>0.79<br>0.91 |
| Three- | dimension | case | $Cl^-$ rotation:<br>α centered<br>$Li^+$ cluster | maximum | maximum<br><br>0.5 |
| α centered<br>$Li^+$ cluster<br>(i) $Cl^-$ motion along <101><br>(ii) $Cl^-$ motion along <110> | minimum<br><br>0.71<br>0.14 | minimum<br><br>0.27<br>0.012 | F centered<br>$Li^+$ cluster | | 1.3 |

* $r_d$ is the magnitude of off-center displacement in units of the anion-to-cation separation of the KCl lattice, $a$ = 3.147 Å. $\delta_{EE}$ is the potential energy barrier for anion motion.

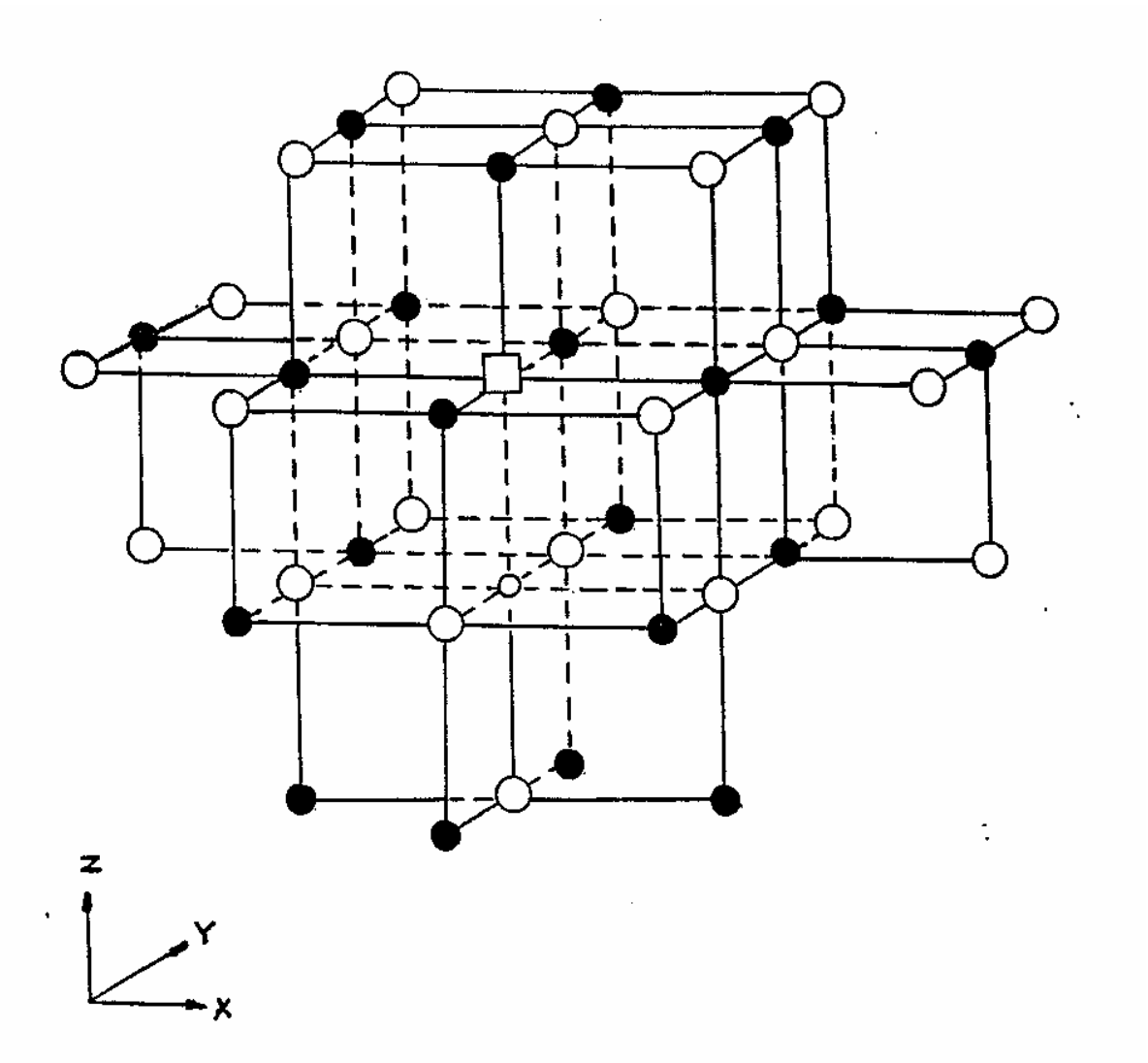

Figure 1: Model of $Li^+$ impurity-containing cluster of *fcc* KCl lattice: We constructed various clusters according to the symmetry of the defects and the direction of $Li^+$ or $Cl^-$ displacement. Symbols: •– $K^+$ ion, °– $Li^+$ impurity ion, O– $Cl^-$ ion, □– $Cl^-$ ion vacancy.

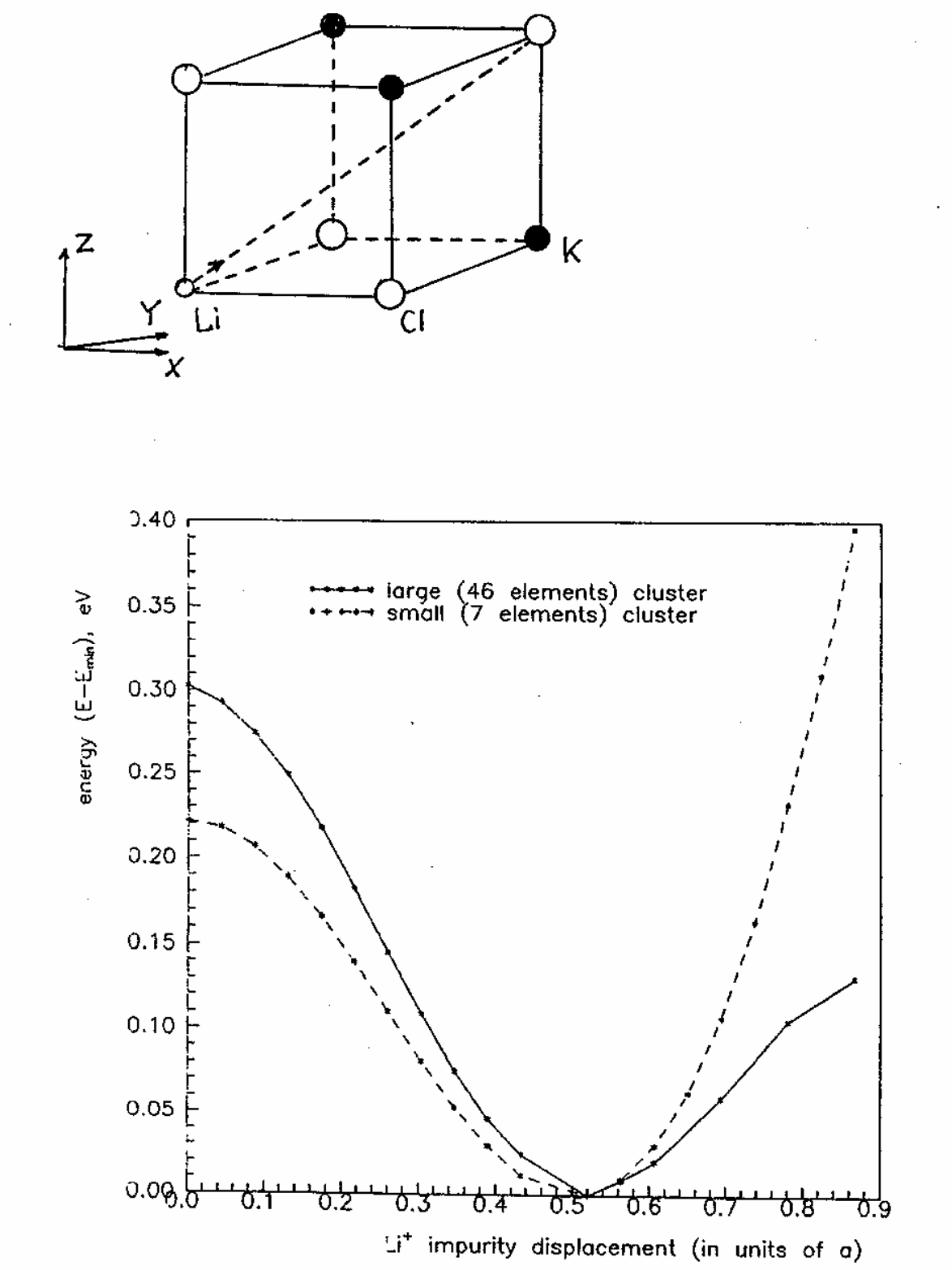

Figure 2: Off-center $Li^+$ impurity motion along <111>: Calculated total total energies E for both large and small clusters against $Li^+$ impurity displacement. a = 3.147 Å is the anion-to-cation separation in KCl.

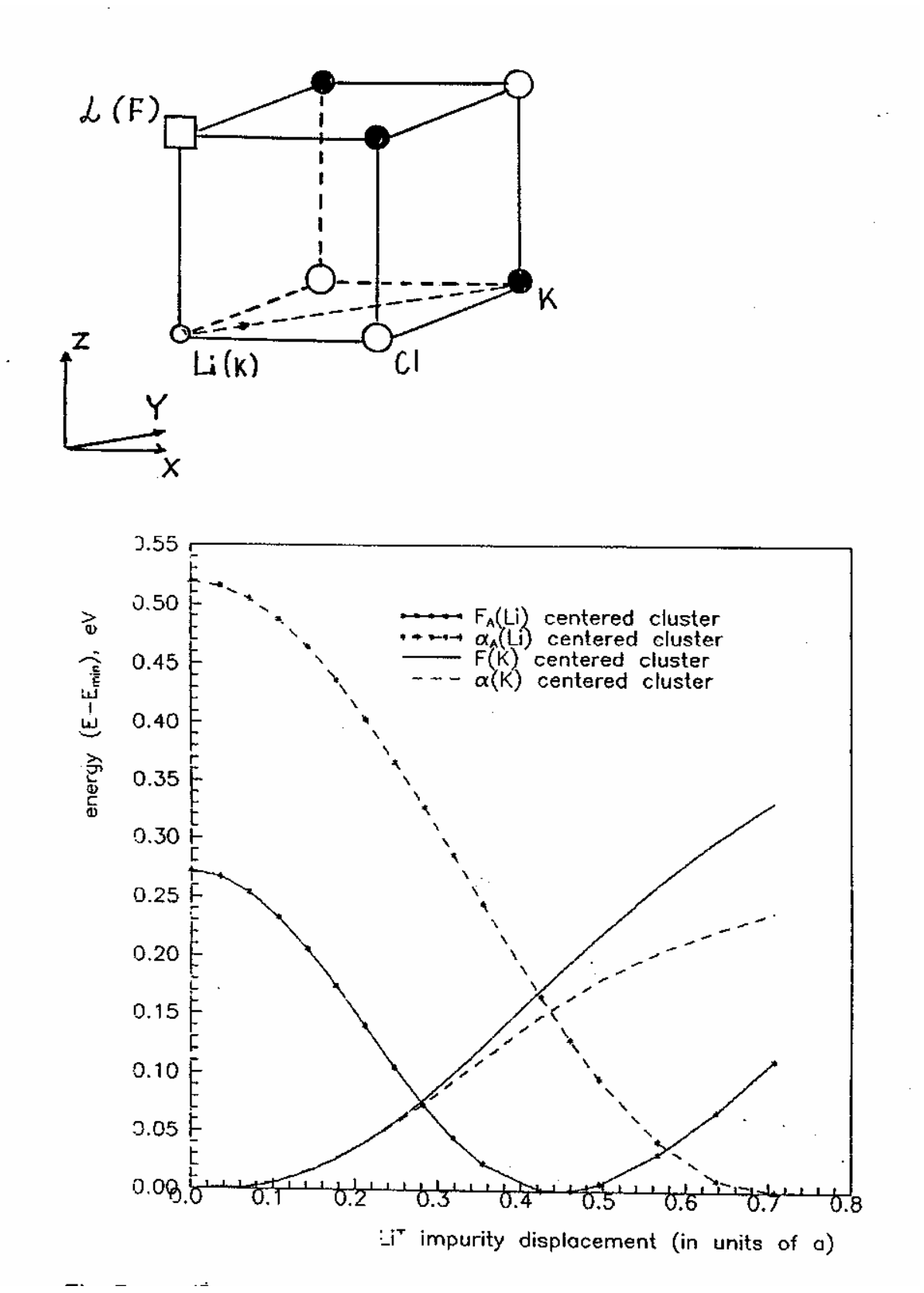

Figure 3: $Li^+$ impurity displacement along <110>: Calculated total energies E for both $F_A$ (Li) and $\alpha_A$ (Li) centered clusters against the $Li^+$ displacement. The lines without symbols are the potential energies of impurity-free clusters with $K^+$ substituting for $Li^+$.

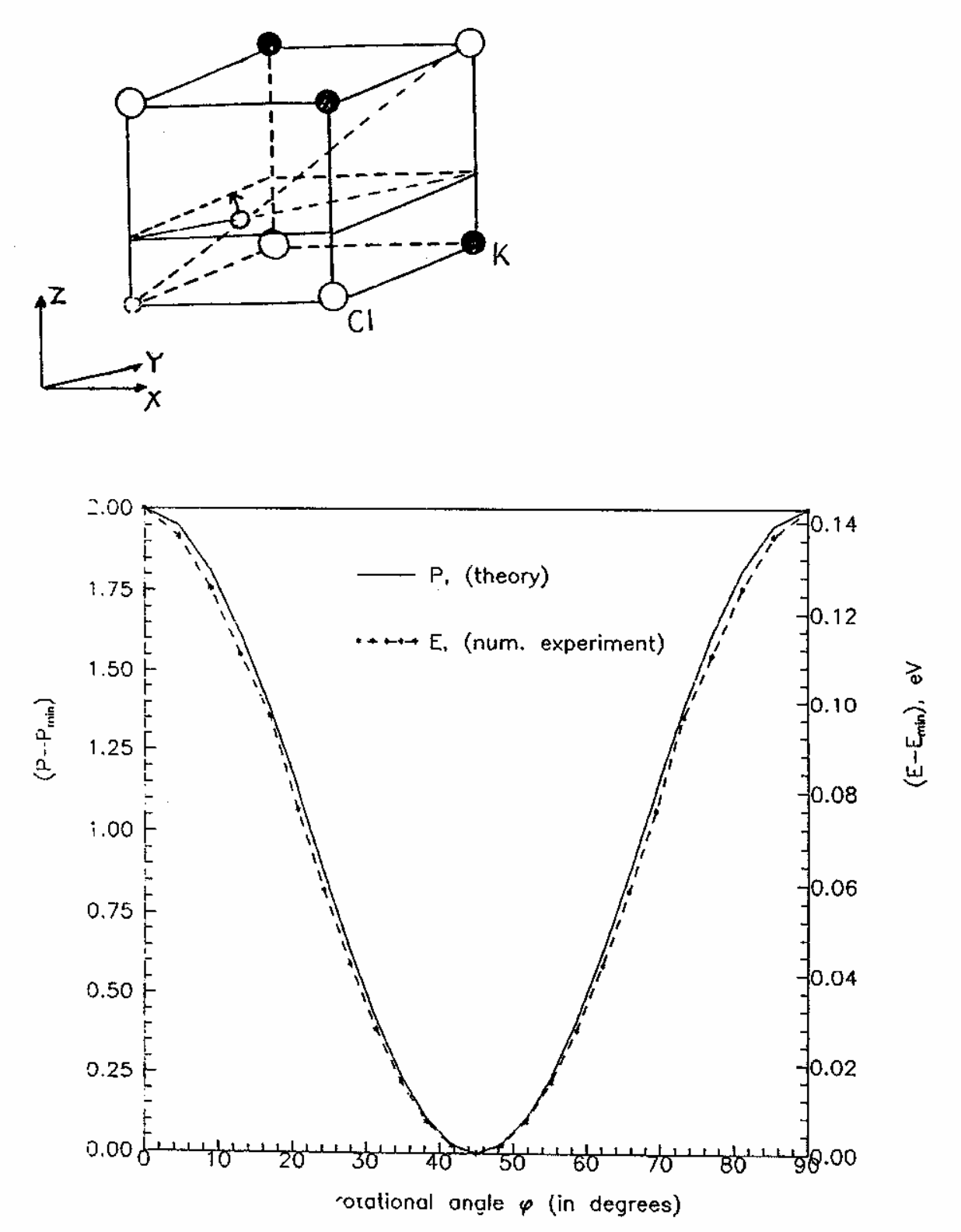

Figure 4: Rotation of off-centered $Li^+$ impurity around <001> in a plane parallel to XY plane: Comparison of the angle-dependent part of the theoretical vibronic potential P = cos(4φ) with numerical calculated total energy E against rotational angle φ.

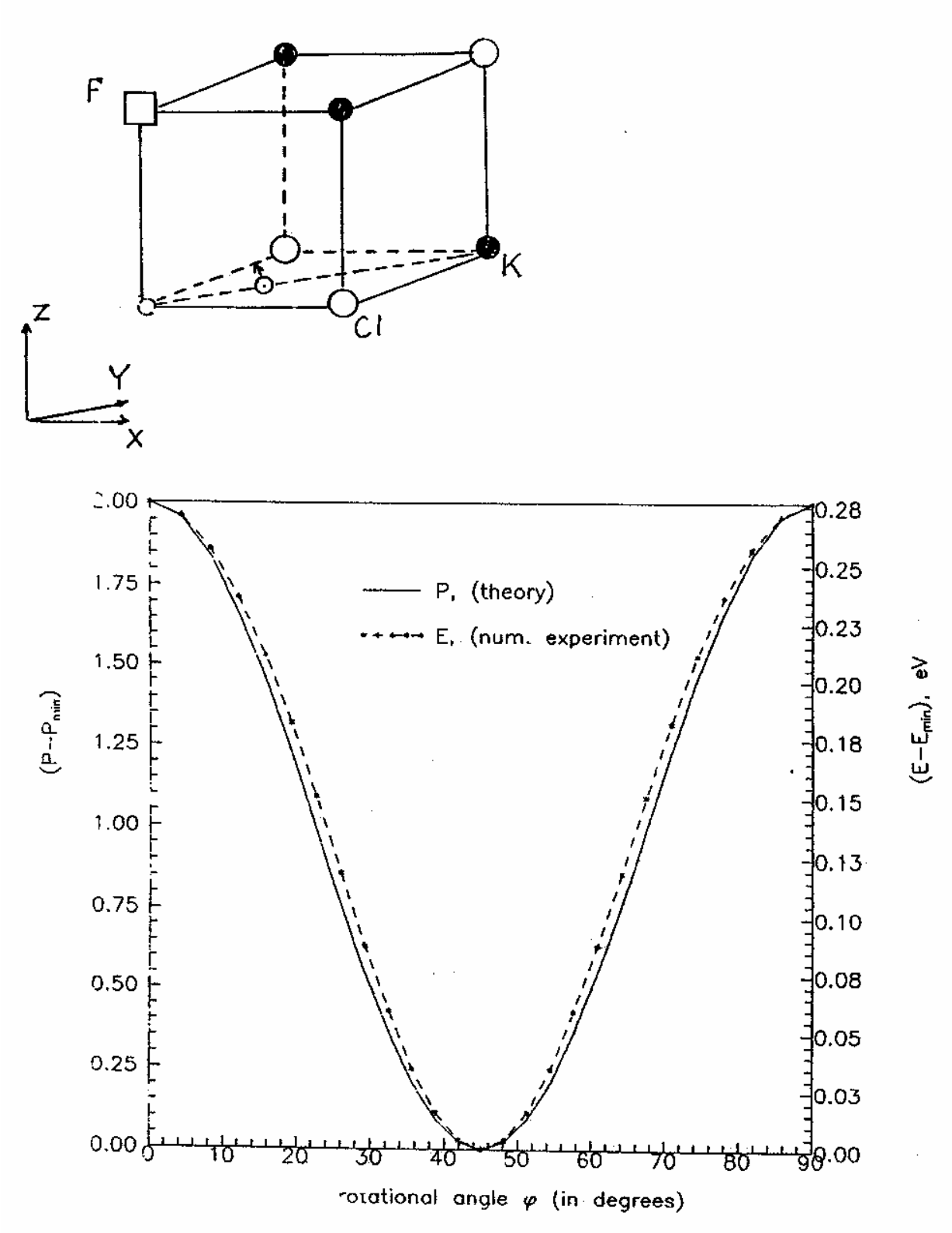

Figure 5: Rotation of off-center $Li^+$ impurity around the on-center site in a XY plane: Comparison of the angle-dependent part of the theoretical vibronic potential $P = \cos(4\varphi)$ with the numerical calculated total energy E against the rotational angle $\varphi$.

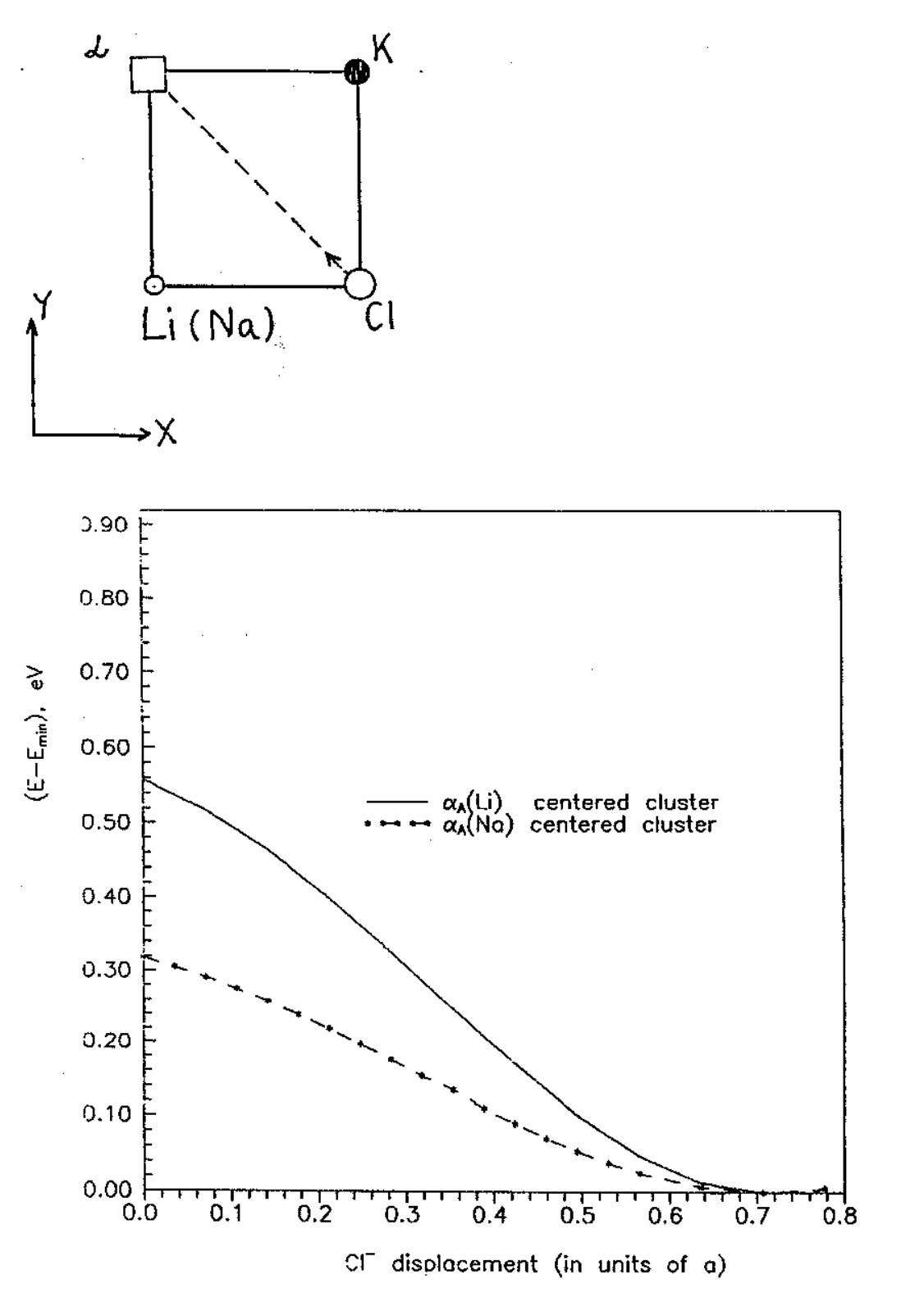

Figure 6: The anion motion along <$\bar{1}$10> towards an α center: Calculated total energies E against $Cl^-$ displacement for $Li^+$ (solid line) or $Na^+$ (dashed line) containing a 24-element cluster.

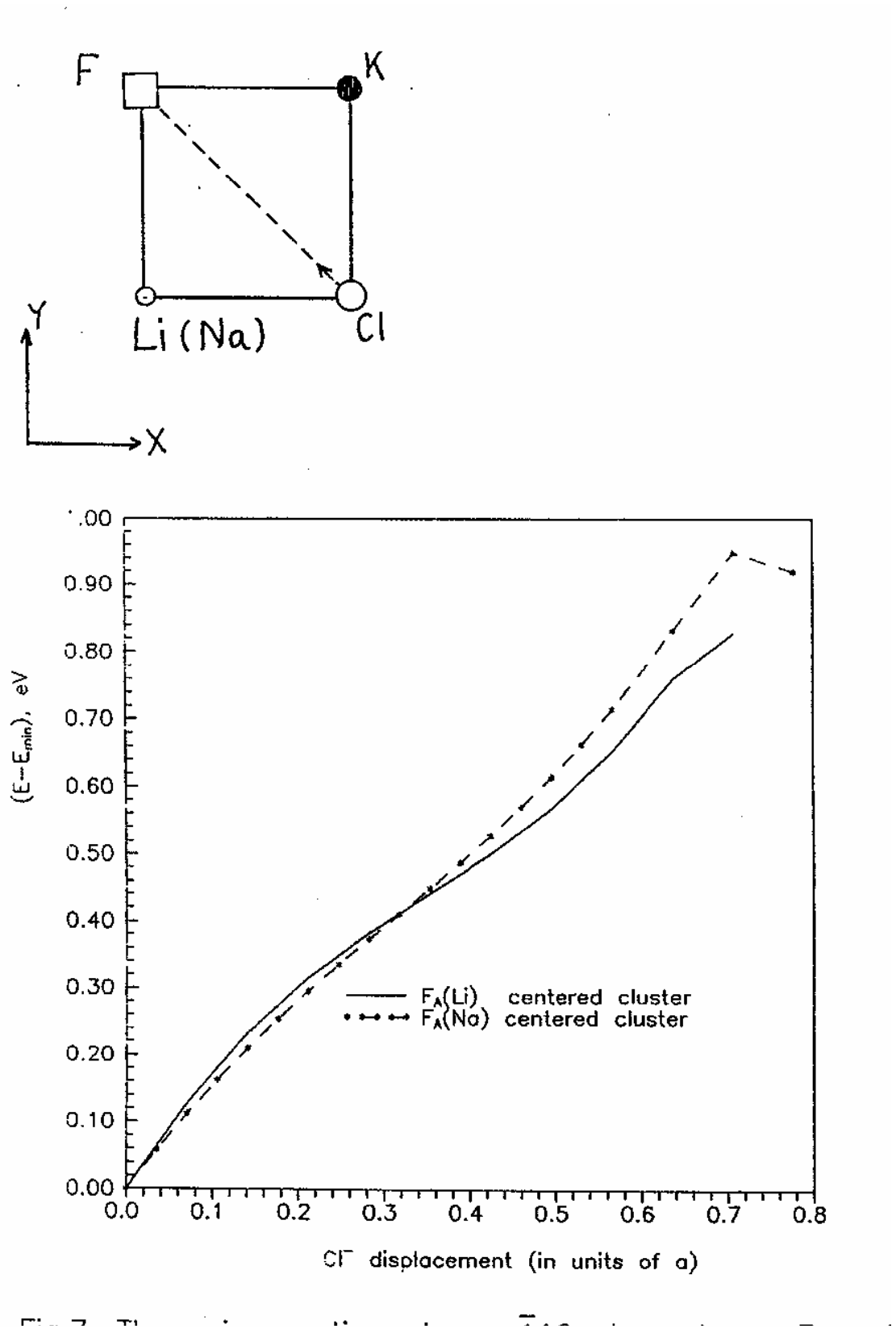

Figure 7: The anion motion along <110> towards an F center: Calculated total energies E against $Cl^-$ displacement for $Li^+$ (solid line) or $Na^+$ (dashed line) containing a 24-element cluster.

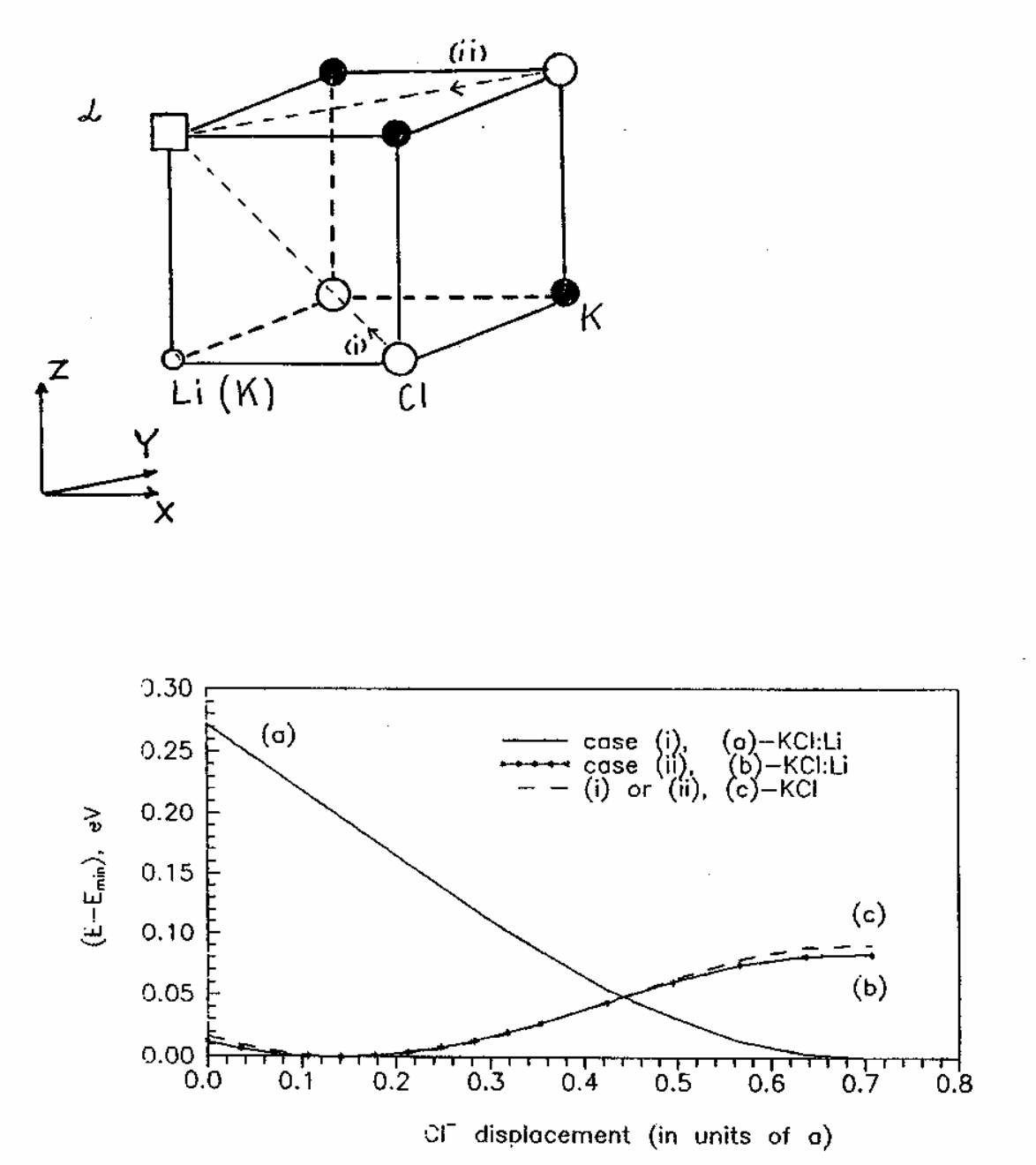

Figure 8: The anion motion towards an α center:
(i)- along <$\underline{1}$01> in XZ plane;
(ii)- along <$\underline{1}$10> in XY plane:
Calculated total energies E against $Cl^-$ displacement in a $Li^+$ containing 46-element cluster (curves (a),(b)) or in impurity-free cluster (curve (c)).

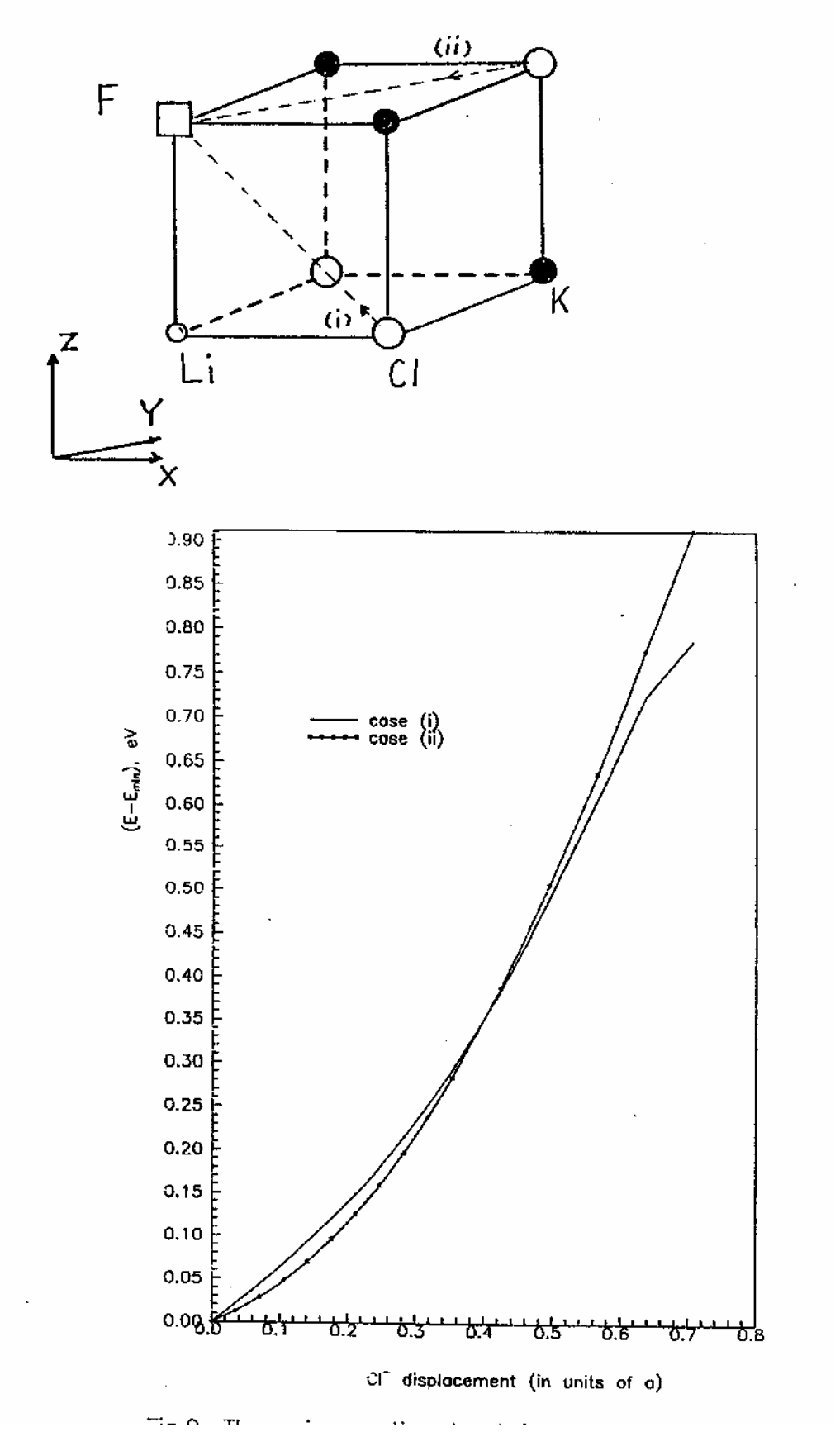

Figure 9: The anion motion towards an F center:
(i)- along <$\underline{1}$01> in XZ plane;
(ii)- along <$\underline{11}$0> in XY plane:
Calculated total energies E against $Cl^-$ displacement in a $Li^+$ containing 46-element cluster.

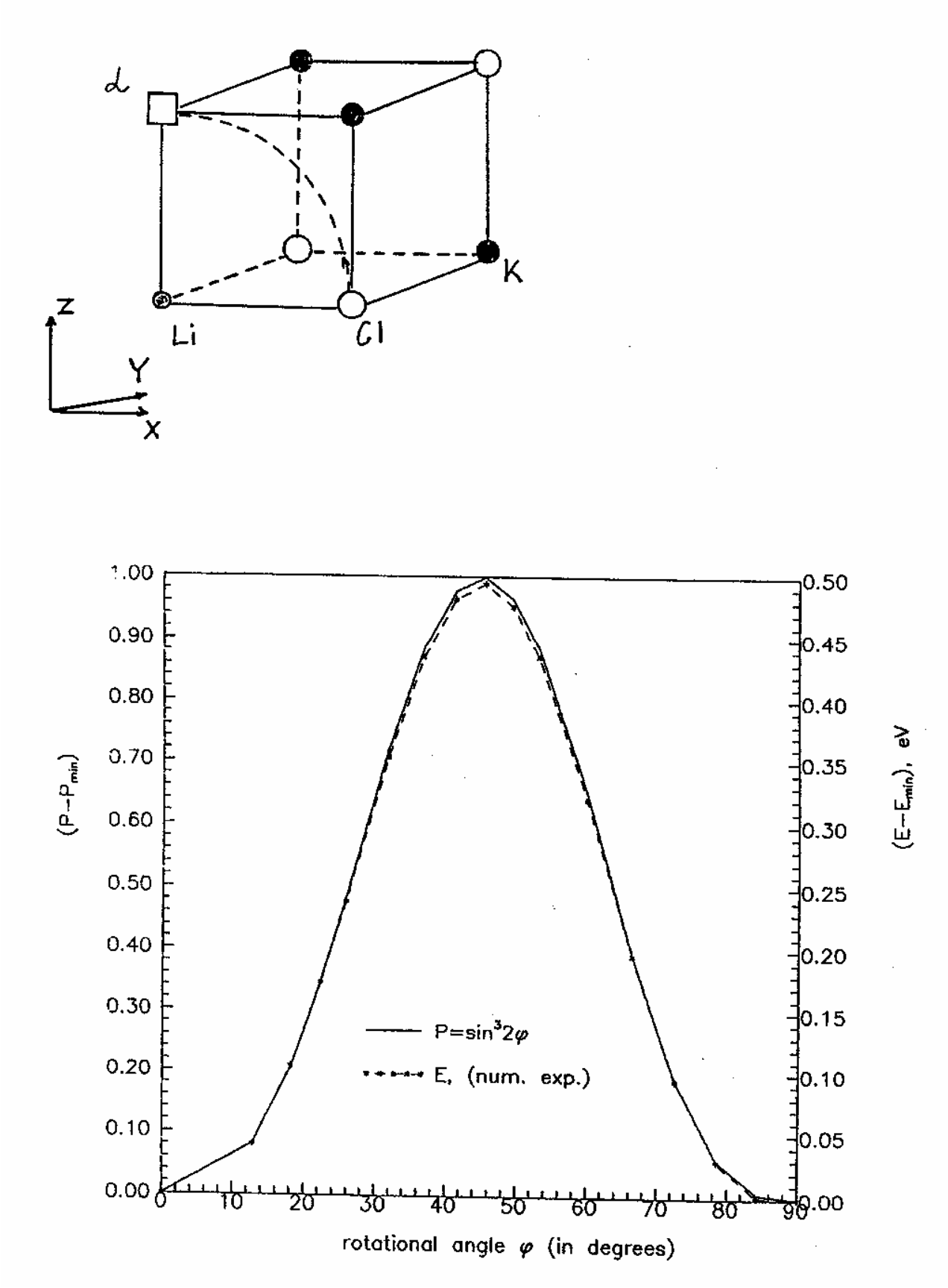

Figure 10: The anion rotation in XZ plane around $Li^{+}$ impurity in a 46-element cluster: Comparison of the numerically calculated total energy E with $P = \sin^3(2\varphi)$ against the rotational angle $\varphi$.

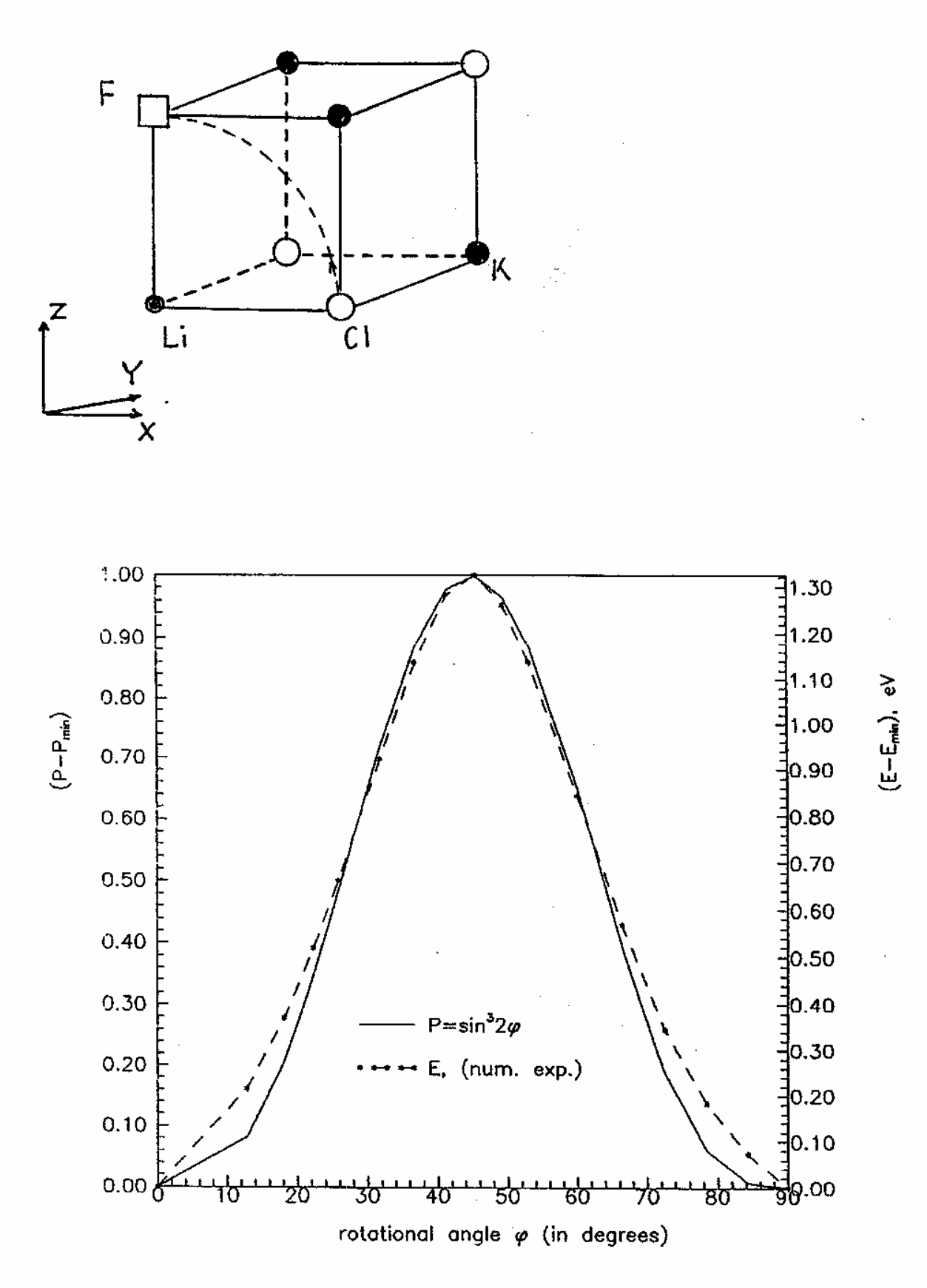

Figure 11: The anion rotation in XZ plane around $Li^+$ impurity in a 46 element cluster: Comparison of the numerically calculated total energy E with $P = \sin^3(2\varphi)$ against the rotational angle $\varphi$.